\documentclass[final,5p,times,twocolumn]{elsarticle}

\usepackage{graphicx}

\usepackage{amssymb}

\def \neq{\(\mathrm{n}_\mathrm{eq}\)/\(\mathrm{cm}^2\) }
\def \stn{S/N\ }

\journal{Nuclear Instruments and Methods A}

\begin{document}

\begin{frontmatter}



\title{Czochralski Silicon as a Detector Material for S-LHC Tracker Volumes}

\author[A]{Leonard Spiegel\corref{*}}
\author[B]{Tobias Barvich}       
\author[C]{Burt Betchart}      
\author[D]{Saptaparna Bhattacharya}
\author[E]{Sandor Czellar}       
\author[C]{Regina Demina}
\author[B]{Alexander Dierlamm}        
\author[B]{Martin Frey}          
\author[C]{Yuri Gotra}         
\author[E]{Jaakko H\"ark\"onen}
\author[B]{Frank Hartmann}      
\author[E]{Ivan Kassamakov}  
\author[C]{Sergey Korjenevski} 
\author[E]{Matti J. Kortelainen} 
\author[E]{Tapio Lamp\'{e}n}    
\author[E]{Panja Luukka}        
\author[E]{Teppo M\"aenp\"a\"a} 
\author[E]{Henri Moilanen}
\author[D]{Meenakshi Narain}
\author[B]{Maike Neuland}
\author[C]{Douglas Orbaker}      
\author[B]{Hans-J\"urgen Simonis}
\author[B]{Pia Steck}     
\author[E]{Eija Tuominen}      
\author[E]{Esa Tuovinen}      
\address[A]{Fermi National Accelerator Laboratory, Batavia, IL, USA} 
\address[B]{Universit\"at Karlsruhe (TH), Institut f\"ur Experimentelle Kernphysik, Karlsruhe, Germany} 
\address[C]{University of Rochester, Department of Physics and Astronomy, Rochester, NY, USA} 
\address[D]{Brown University, Providence, RI, USA}
\address[E]{Helsinki Institute of Physics, Helsinki, Finland} 

\cortext[*] {Corresponding author. Address: FNAL, Batavia, IL, USA, Tel +1 630 840 2809, E-mail lenny@fnal.gov}

\begin{abstract}

With an expected ten-fold increase in luminosity in S-LHC, the
radiation environment in the tracker volumes will be considerably
harsher for silicon-based detectors than the already harsh LHC
environment.  Since 2006, a group of CMS institutes, using a modified
CMS DAQ system, has been exploring the use of Magnetic Czochralski
silicon as a detector element for the strip tracker layers in S-LHC
experiments. Both p+/n-/n+ and n+/p-/p+ sensors have been
characterized, irradiated with proton and neutron sources, assembled
into modules, and tested in a CERN beamline. There have been three
beam studies to date and results from these suggest that both p+/n-/n+
and n+/p-/p+ Magnetic Czochralski silicon are sufficiently radiation
hard for the $R>25$ cm regions of S-LHC tracker volumes.  The group
has also explored the use of forward biasing for heavily irradiated
detectors, and although this mode requires sensor temperatures less
than -50\,$^\circ$C, the charge collection efficiency appears to be
promising.

\end{abstract}


\end{frontmatter}


\section{Introduction}

Although the time-scale for Large Hadron Collider (LHC) luminosity
upgrades has recently slipped, the ultimate goal of increasing the
design LHC design luminosity by a factor of ten remains
unchanged. This increase means that trackers planned for S-LHC
experiments will require sensor materials in the innermost layers with
an order of magnitude improvement in radiation hardness over what is
presently in use. For pixel systems, which typically extend from just
outside the beam pipe to a radius of about 25~cm, the sensor material
should be able to survive fluences in excess of $1\times 10^{16}$
1~MeV neutron equivalents per square centimeter.\footnote{For brevity
the fluence unit is abbreviated as \neq in this note.} However, pixel
detectors are relatively small in total sensor area, so that expensive
semiconductor materials, such as diamond, are not out of the
question. Pixel systems with their close proximity to the beam pipe
also offer the possibility of a periodic replacement strategy.

In the strip tracker regions, which start at around $R=25$~cm, the
maximum fluences are significantly reduced, to about $1\times
10^{15}$~\neq.  However, LHC strip trackers tend to have large sensor
areas---some 200~m$^2$ in the case of CMS---and do not provide the
option of easy replacement.  It is also assumed that for ease of
construction and operation, experiments will want to use a single
technology for the strip sensors.

Since 2006 a group of CMS institutes (SiBT group \cite{ref:group}) has
been studying, based on the use of a modified CMS DAQ system, the use
of Magnetic Czochralski (MCz) silicon as a potential replacement for
the Float Zone (FZ) silicon that is presently used in the CMS strip
tracker.  The Czochralski process utilizes a quartz crucible for
containing the molten silicon, which facilitates the production of
large area wafers but at the same time is a source of
contaminants. However, it has been shown that one of contaminants,
oxygen, actually improves the radiation hardness of the silicon
\cite{ref:oxy1,ref:oxy2}. By carrying out the Czochralski process in
the presence of of a magnetic field the concentration of oxygen can be
precisely controlled.

\section{Beam Telescope}

\begin{figure}[hbt] 
\centering 
\includegraphics[width=0.5\textwidth,keepaspectratio]{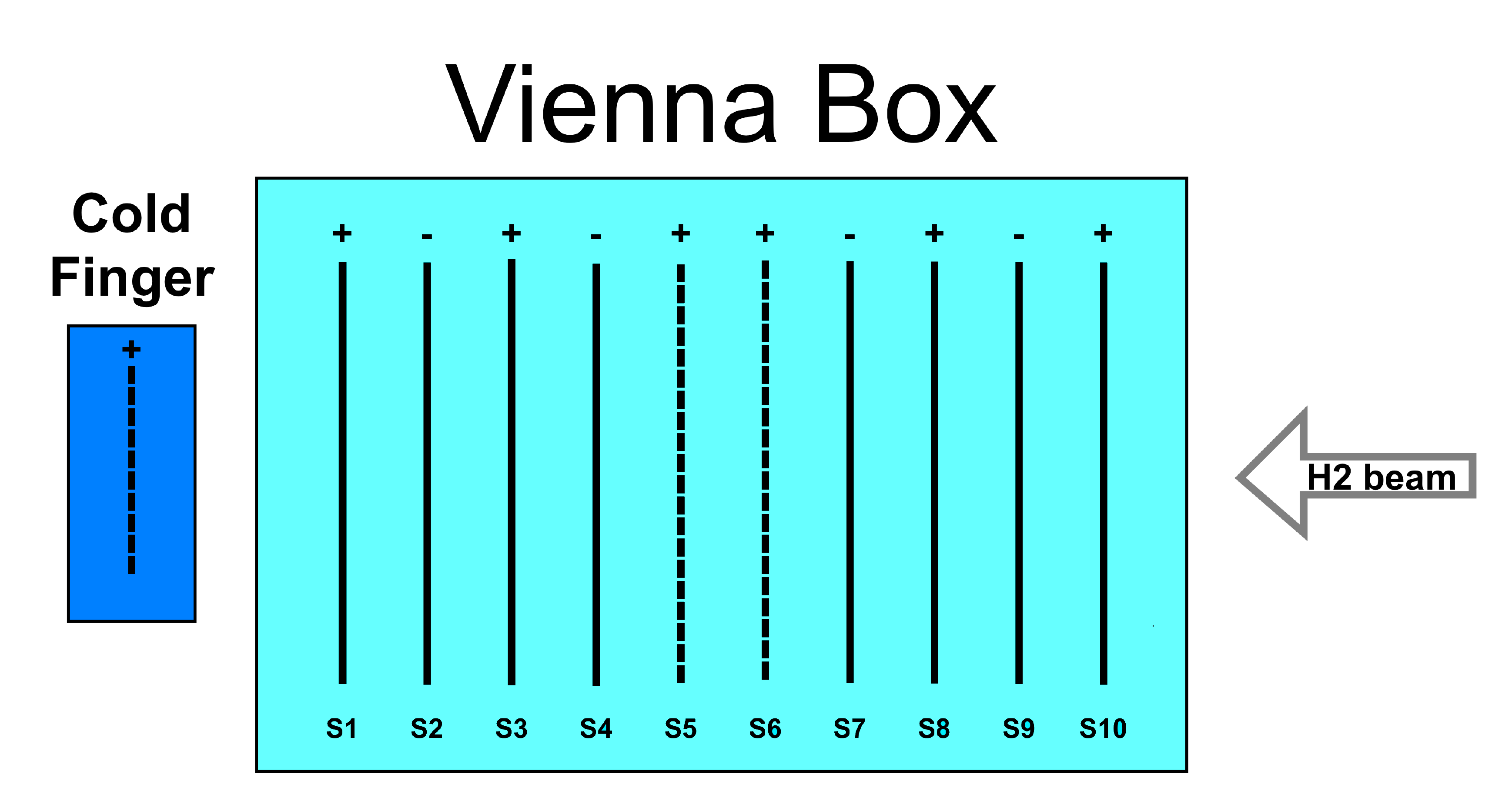}
\caption{Beam telescope used for MCz studies.}
\label{fig:telescope}
\end{figure}

The beam telescope used by the SiBT group has been previously
documented \cite{ref:sibt}. It is based on the cold box developed by
the Vienna HEP group for long-term testing of CMS silicon strip
modules during the construction period. CMS DAQ components and
software are used for the readout of modules containing CMS front-end
hybrids. Figure~\ref{fig:telescope} shows a schematic of the 10-slot
Vienna box and its orientation relative to the CERN SPS H2 beam
(225~GeV/$c$ muons or pions). There is a 4~cm spacing between slots
and the box has an internal clearance height of 17~cm and a depth of
about 44~cm. The box is capable of achieving an inside temperature
of -25\,$^{\circ}$C with detectors under power in all 10 slots;
was run as close to this as possible when irradiated detectors 
were present so as to limit the leakage currents.

\begin{figure}[hbt] 
\centering 
\includegraphics[width=0.5\textwidth,keepaspectratio]{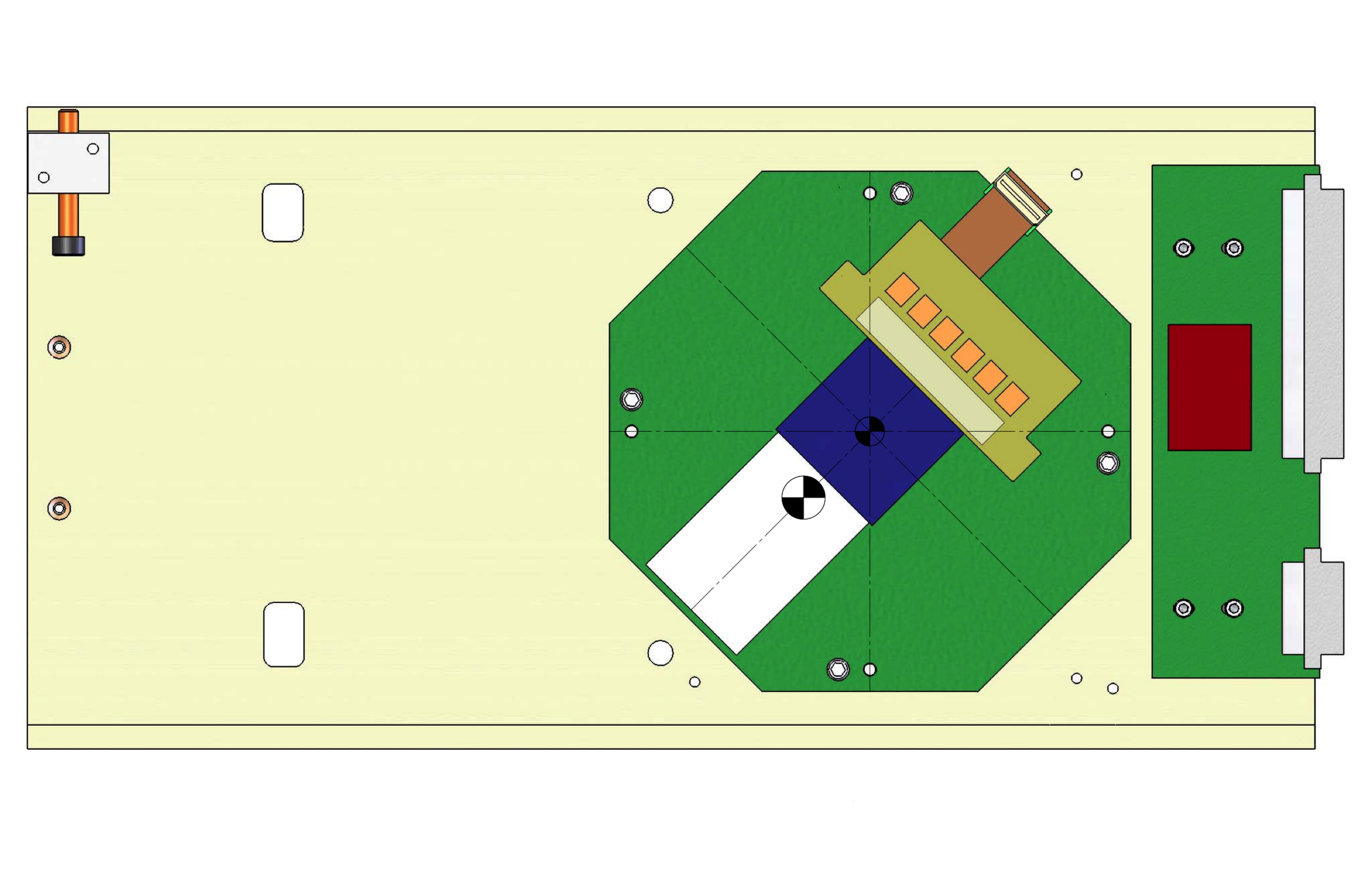}
\caption{Vienna box insertion plate illustrating the $+45^\circ$
orientation for a reference module (long, white sensor) or a DUT
(short, blue sensor).}
\label{fig:brass}
\end{figure}

The reference planes, which normally occupy slots 1-4 and 7-10, are
based on 4$\times$10~cm$^2$ Hamamatsu \cite{ref:hpk} sensors provided
by the D0 collaboration and leftover CMS outer barrel
hybrids. Although the reference detectors have a 60~$\mu$m pitch, they
also contain intermediate strips so that the resolution of planes is
more consistent with that of a 30~$\mu$m pitch detector (and in fact
has been measured to be as small as 6.5~$\mu$m). Within the Vienna box
the reference planes are mounted either in $+45^\circ$ or $-45^\circ$
orientations as shown in Fig.~\ref{fig:brass}.  With this geometry the
uncertainty in beam track positions in slots 5 and 6, as determined by
the four ``+'' reference planes, is around 3.3~$\mu$m. Details on the
calibration and alignment of reference planes can be found in
Ref. \cite{ref:cal}.

Figure 1 also shows a second, single-detector cold enclosure, which
was installed immediately downstream of the Vienna box. This Cold
Finger used a three-stage Peltier element for cooling, but with a smaller
volume and tighter insulation was able to achieve an internal
temperature as low as -53\,$^\circ$C. Detectors installed in this box had
a ``+'' orientation and were operated in both reverse and forward bias
modes.

\section{Detectors Under Test}

The SiBT group has studied a number of MCz detectors starting with the
initial beam study in the summer of 2007. Table~\ref{tab:duts} lists
those detectors that are the subject of this note. In the table and
subsequent sections ``n(-type)'' refers to p+/n-/n+ sensors and
``p(-type)'' to n+/p-/p+ sensors.  Within the Vienna box the detectors
under test (DUT) were installed in slots 5 or 6 given the minimized
beam track uncertainty in these locations as described in the previous
section.

The MCz sensors used in the DUT modules were processed at the Helsinki
University of Technology Centre for Micro and Nanotechnology
(Micronova) facility. The raw material was provided by Okmetic 
\cite{ref:okm} in the form of 4 inch wafers with a thickness of
300~$\mu$m and a nominal resistivity of 900 $\Omega$cm. Details on the
processing can be found in Ref. \cite{ref:mcz}. With an area of
4.1$\times$4.1~cm$^2$ the MCz sensors are considerably shorter than
the reference sensors, as is illustrated in Fig. 2. All of the sensors
had a strip pitch of 50~$\mu$m and a strip width of 10~$\mu$m.

Similarly, sensors were also fabricated from Topsil \cite{ref:top} FZ
wafers, obtained from an RD50 \cite{ref:rd50} common order, using the
same masks from the MCz production. The Topsil wafers had a thickness
of 285~$\mu$m.

Non-irradiated MCz sensors fully deplete at around 350 volts and
non-irradiated Topsil FZ sensors at around 10 volts.

The MCz and FZ sensors were irradiated, prior to module assembly, with
25~MeV protons at the Universit\"at Karlsruhe and, in a limited number
of cases, also with 3-45 MeV neutrons at the Universit\'e Catholique
de Louvain.  Irradiation levels for the detectors covered in this note
are listed in Table 1. As described in Section 2 the Vienna box was
operated close to the coldest setting for the irradiated detectors.
Actual sensor temperature were somewhat higher, though, given the
power consumption of the CMS hybrids and the leakage currents.

\begin{table}[h]\centering
  \caption{Detectors Under Test}
  \begin{tabular}{ccclcc}
    year & fluence & proton \% & sensor & box & bias \\
    \hline
    2007  & non-irrad.           &     &  nMCz & Vienna & rev \\ 
    2008  & non-irrad.           &     &  nFZ  & Vienna & rev \\ 
    2009  & $1.0\times 10^{14}$  & 100 &  nFZ  & Vienna & rev \\
    2008  & $2.2\times 10^{14}$  & 100 &  nFZ  & Vienna & rev \\
    2009  & $3.0\times 10^{14}$  & 100 &  nFZ  & Vienna & rev \\
    2008  & $6.1\times 10^{14}$  &  84 &  nMCz & Vienna & rev \\
    2008  & $1.1\times 10^{15}$  &  91 &  nMCz & Vienna & rev \\
    2008  & $1.6\times 10^{15}$  &  94 &  nMCz & Vienna & rev \\
    2009  & $2.0\times 10^{15}$  & 100 &  pMCz & CF     & rev/for \\
    2008  & $2.8\times 10^{15}$  & 100 &  nMCz & CF     & rev/for \\
    2009  & $3.1\times 10^{15}$  & 100 &  nMCz & Vienna & rev/for \\
    2009  & $4.9\times 10^{15}$  & 100 &  nMCz & CF     & rev/for \\
  \end{tabular}
  \label{tab:duts}
\end{table}

\section{Charge Collection Efficiency}

Charge clusters in the DUTs are determined in a standard way: seed
strips are identified based on the requirement that the signal to
noise ratio (\stn) exceeds 2.5. Neighboring strips are then added to
the seed if their \stn exceeds 2.0 and the final cluster must have a
\stn greater than 2.75. The total charge associated with a cluster is
then taken to be the sum of the charges of the two largest adjacent
strips within the cluster.  As the DUTs are normal to the beam, the
DUT clusters tend to be dominated by single-strip clusters. In these
cases the cluster charge is simply the charge of the single strips.
Clusters are classified as either ``on-track'' or ``off-track''
depending on their proximity to the beam track.

In some cases where strip and cluster charges are likely to be below
threshold a {\it non-clustering} algorithm \cite{ref:ns} has been
used. In this approach the cluster charge is taken to be the sum of
charges for the two strips that are closest to the intersection of the
beam track, as determined by the reference planes, and the DUT.

\begin{figure}[hbt] 
\centering 
\includegraphics[width=0.5\textwidth,keepaspectratio]{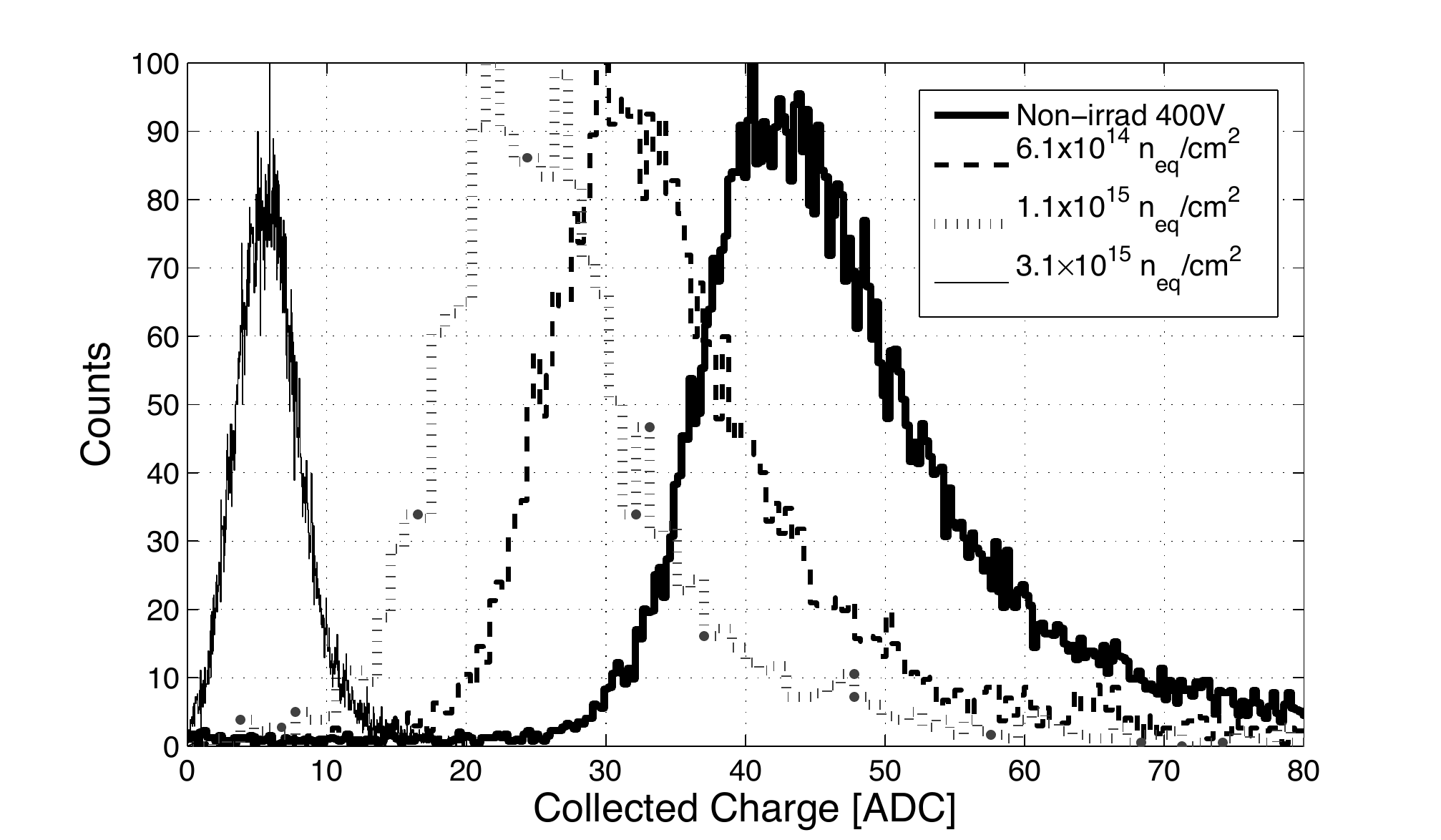}
\caption{Superimposed cluster charge distributions for four of the
n-type MCz detectors. The distributions have been scaled so that there
are 100 counts in the highest bin and they are all based on the
non-clustering method described in the text. Bias voltage of 600~V
and 420~V were applied for the $1.1\times 10^{15}$~\neq and
$3.1\times 10^{15}$~\neq detectors, respectively, and these are
well bellow the full depletion values.}

\label{fig:landau}
\end{figure}

The signal associated with a particular bias voltage is taken to be
the most probable value of a Landau function convoluted with a
Gaussian function that has been fit to the on-track
data. Figure~\ref{fig:landau} shows the distribution of cluster
charges for four of the n-type MCz detectors.  A MPV of about 40~ADC
counts for the non-irradiated MCz sensor is in agreement with the
level seen in the reference detectors. As can be seen in the figure,
increasing fluence levels result in correspondingly lower charge
collection efficiencies. It should be pointed out that 600~V was the
maximum bias voltage\footnote{600~V is the approximate safe operating
limit for high voltage cables presently installed in CMS; it is
unlikely that these cables will be changed for S-LHC.} that was
applied during the studies.  At 600~V the $6.1\times 10^{14}$~\neq is
just below full depletion whereas the $1.1\times 10^{15}$~\neq and
$3.1\times 10^{15}$~\neq detectors are well below their full depletion
values. However, the $1.1\times 10^{15}$~\neq detector does exhibit a
MPV of about 20~ADC at this voltage, corresponding to a charge
collection efficiency of 50\%. Coupled with a noise measurement of
2~ADC counts \cite{ref:res} this leads to a predicted S/N$>$10 at the
end of S-LHC running.

\begin{figure}[hbt] 
\centering 
\includegraphics[width=0.5\textwidth,keepaspectratio]{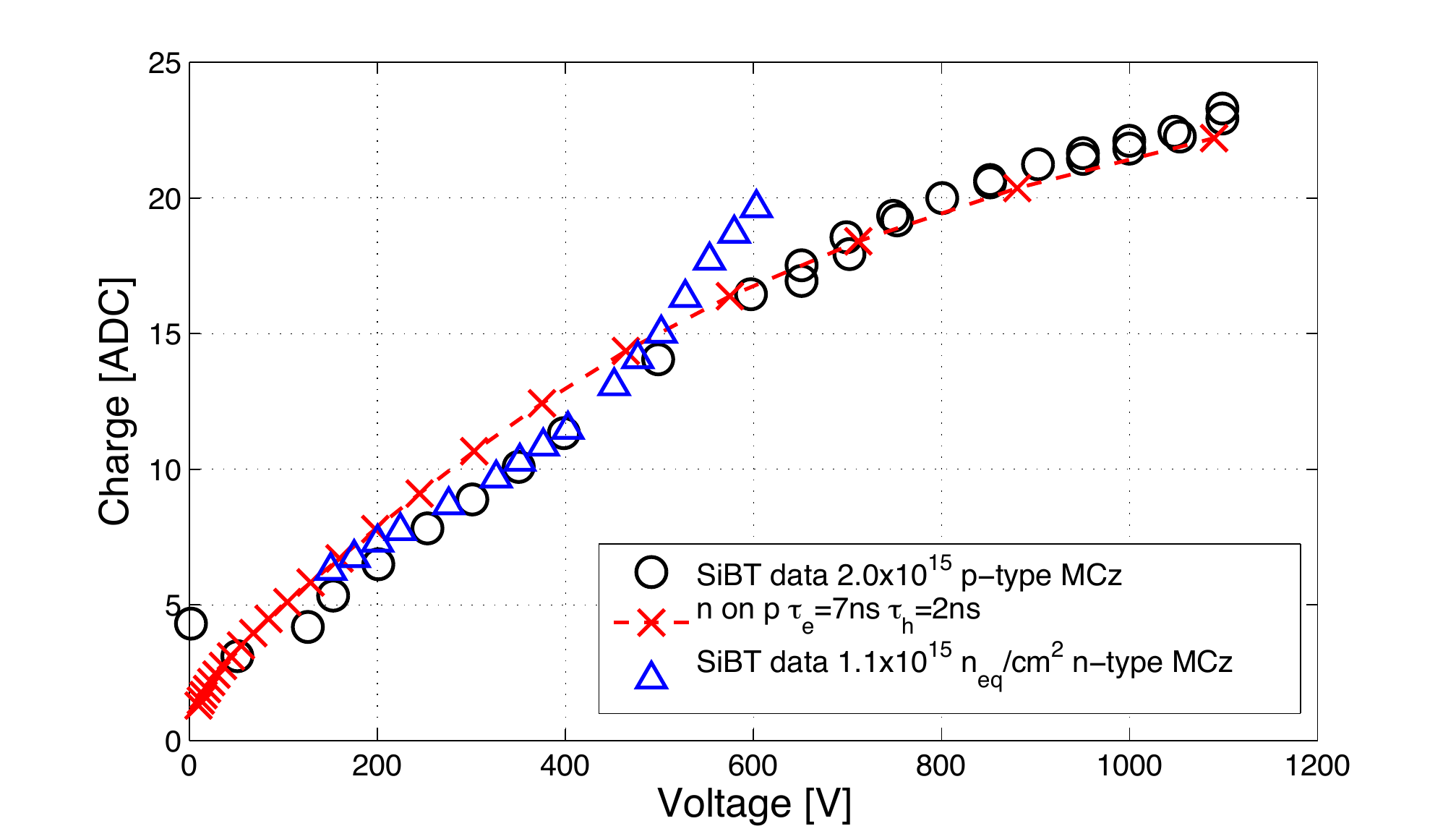}
\caption{Comparison of n-type and p-type MCz charge collection efficiency. 
The dashed line represents a transport model prediction that has been tuned
to fit the p-type data. A non-clustering approach was used for the 
the p-type data.}
\label{fig:np_comp}
\end{figure}

One p-type MCz detector, irradiated to a fluence of $2.0\times
10^{15}$~\neq, was operated in reverse bias mode in the Cold Finger in
2009.  Figure~\ref{fig:np_comp} compares the signal versus voltage for
the p-type detector with that of the $1.1\times 10^{15}$~\neq n-type
detector.  Despite the different fluence levels the two detectors have
approximately the same charge collection efficiency, at least up to
the maximum 600~V applied to the n-type detector. Also shown on the
figure is the result of a transport model for the p-type detector that
has been tuned to roughly fit the data (by adjusting trapping time
constants for holes and electrons within reasonable ranges).

There is no evidence for avalanche or charge multiplication effects in
any of the SiBT data although $5\times 10^{15}$~\neq and 1000~V is
the largest combination of fluence level and bias voltage that has
been studied to date.

\begin{figure}[hbt] 
\centering 
\includegraphics[width=0.5\textwidth,keepaspectratio]{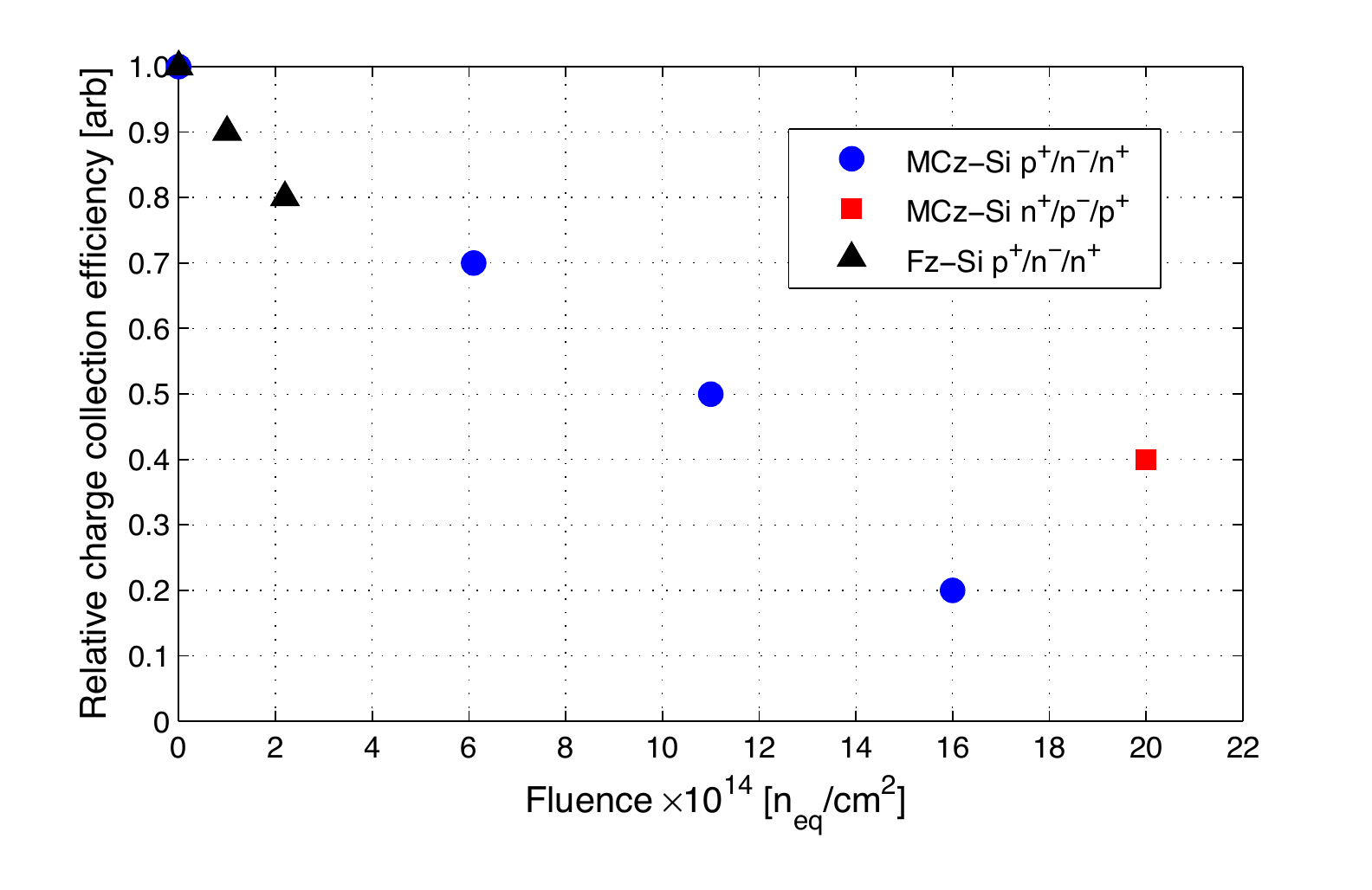}
\caption{Charge collection efficiencies for n-type MCz, p-type MCz,
and (n-type) FZ detectors as a function of fluence. The MCz detectors
have been normalized to the non-irradiated MCz detector and the FZ
detectors to the non-irradiated FZ detector. The last three points 
were taken at 600~V, which is well below the estimated full depletion
voltages.}
\label{fig:fluence}
\end{figure}

Figure~\ref{fig:fluence} shows the charge collection efficiency for
four n-type MCz detectors, one p-type MCz detector, and three n-type
FZ detectors as a function of fluence. The last three points were
measured at 600~V, which is well below the expected full depletion
voltages.

\section{Current Injection}

For heavily irradiated detectors there exists the possibility of
filling trapping sites under forward bias with either holes or
electrons and, assuming the trapping can be continually balanced by
detrapping, establishing a stable electric field throughout the entire
bulk \cite{ref:cid1,ref:cid2}.  The field, which increases toward the
backplane of the detectors as the square-root of the distance from the
injecting junction, is relatively insensitive to the applied voltage
and fluence level. Since detrapping time constants depend
exponentially on temperature, creating a steady-state condition
requires a sensor temperature that is considerably lower than the
optimal -10\,$^\circ$C for irradiated detectors under reverse bias.

Current Injected Detectors (CID) are based on the concept of trapping
and detrapping in forward bias and the SiBT group has explored this
operating mode in parallel with the standard MCz studies
\cite{ref:cid3}. As described in the Beam Telescope section, the Cold
Finger was used for these studies. With an operating temperature as
low as -53\,$^\circ$C the enclosure was capable of providing the
required short detrapping times.

\begin{figure}[hbt] 
\centering 
\includegraphics[width=0.5\textwidth,keepaspectratio]{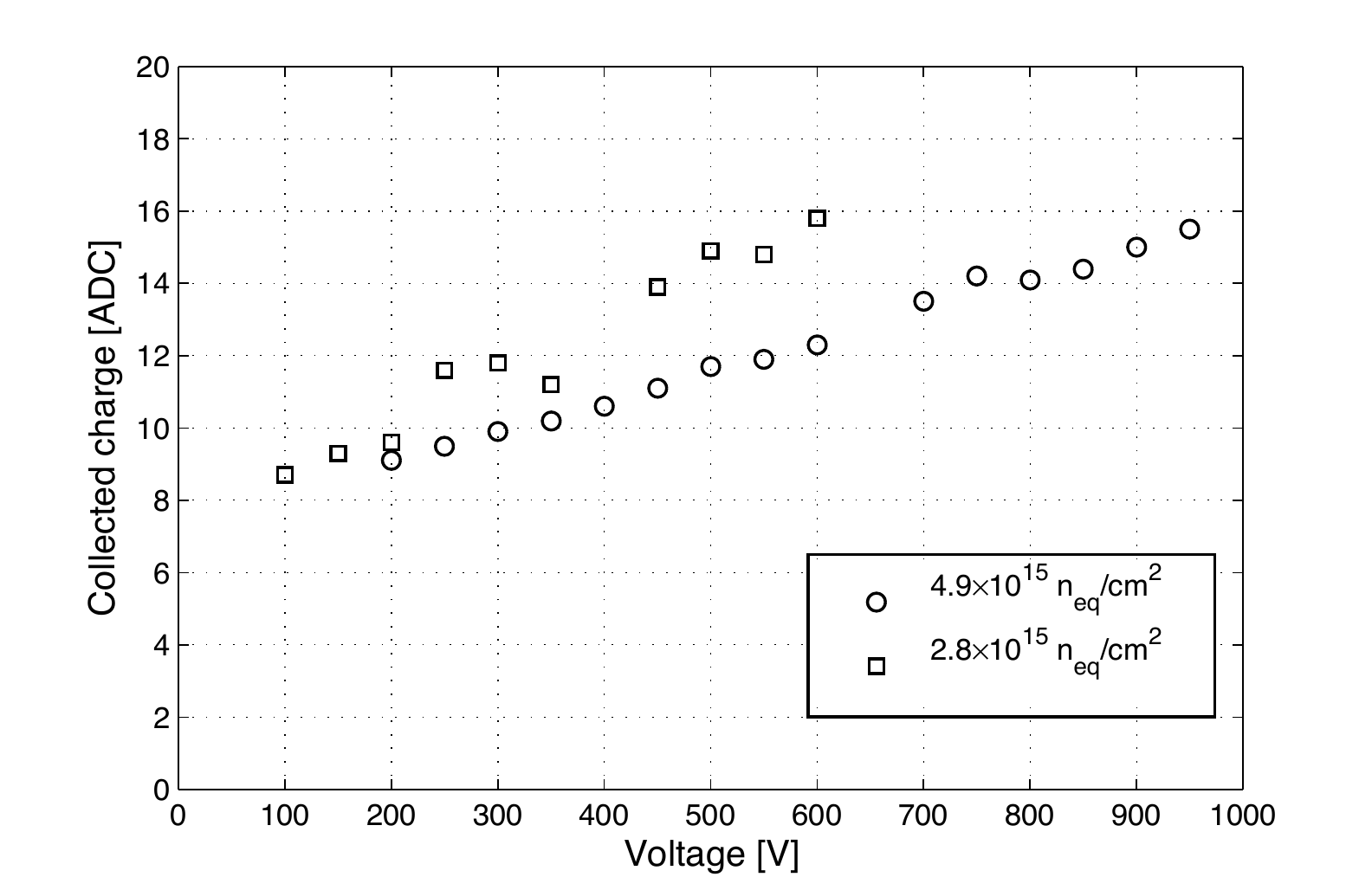}
\caption{Signal versus bias voltage for two heavily irradiated
detectors under forward bias (current injection mode).}
\label{fig:cid}
\end{figure}

Figure~\ref{fig:cid} shows the signal versus bias voltage for two
n-type MCz detectors that have been irradiated to fluences of
$2.8\times 10^{15}$~\neq and $4.9\times 10^{15}$~\neq
respectively. Both detectors were operated in the Cold Finger at the
lowest temperature setting. The $2.8\times 10^{15}$~\neq detector has
a noise value around 2~ADC counts at 600~V, which would imply a \stn
around 8. Although this level is probably at the margin of
acceptability, the current injection mode might provide a means for
extending the lifetime of the innermost layers in S-LHC strip
trackers. However, engineering a cooling system capable of providing
sensor temperature less than -50\,$^\circ$C would be a formidable
challenge.

\section{Conclusions and Future Plans}

The beam studies of irradiated MCz detectors suggest that both n-type
and p-type sensors will survive S-LHC integrated fluence in strip
tracker volumes of planned detectors. The p-type detector does have a
higher charge collection than the n-type, which is not unexpected
given the known advantages in collecting electrons over
holes. However, given the extra processing required to isolate the n+
implants in p-type sensors, this may weigh in favor of n-type sensors
for the very large silicon surface areas in the planned strip
trackers.

The CMS Collaboration has recently contracted with Hamamatsu for over
one hundred 6 inch wafers including the following substrates and
thicknesses: MCz 200~$\mu$m, FZ 200~$\mu$m, FZ 100~$\mu$m, epi
100~$\mu$m, and epi 75~$\mu$m.  These will include both n-type and
p-type and for the p-type detectors both p-spray and p-stop isolation
methods are specified. Detector geometries include pixel, long pixel,
and strips.  Once the sensors have been delivered and probed, a large
fraction will be irradiated with proton, neutron, and mixed proton and
neutron sources. A number of the irradiated strip-geometry sensors
will be assembled into modules using CMS hybrids, and the SiBT group
expects to test many of these in a beam study planned for the fall of
2010.

\section{Acknowledgments}

This work has been supported by the Academy of Finland, Fermi Research
Alliance, LLC under Contract No. DEAC02-07CH11359 with the United States
Department of Energy, and the Initiative and Networking Fund of the Helmholtz 
Association, contract HA-101 (``Physics at the Terascale''). The results 
have been obtained in the frameworks of CERN CMS, RD50, and RD39
collaborations.




\end{document}